\documentclass[%
 reprint,
 amsmath,amssymb,groupedaddress,superscriptaddress,
 aps,longbibliography,floatfix,preprint,onecolumn
]{revtex4-2}
\usepackage[T1]{fontenc}
\usepackage{graphicx}
\usepackage{dcolumn}
\usepackage{bm}
\usepackage{amsmath}
\usepackage{lmodern}
\usepackage{mathtools}
\usepackage[colorlinks,citecolor=black,linkcolor=black,urlcolor=black]{hyperref}

\usepackage[normalem]{ulem} 
\usepackage{upgreek}

\hypersetup{
    colorlinks,
    linkcolor= [rgb]{0.1804,0.1882,0.5725},
    citecolor=[rgb]{0.1804,0.1882,0.5725},
    urlcolor=[rgb]{0.1804,0.1882,0.5725}
}
\usepackage{chemformula}

\setcitestyle{super}

\begin{document}
\title{ Old experiments in new light: Young's double-slit and Stern-Gerlach experiments in liquid crystal microcavities}

\author{Mateusz\,Kr\'ol}
\thanks{These authors contributed equally to this work}
\author{Katarzyna\,Rechci\'nska}
\thanks{These authors contributed equally to this work}
\affiliation{Institute of Experimental Physics, Faculty of Physics, University of Warsaw, Poland}
\author{Helgi Sigurdsson}
\affiliation{Skolkovo Institute of Science and Technology, Bolshoy Boulevard 30, bld. 1, Moscow, 121205, Russia}
\affiliation{Department of Physics and Astronomy, University of Southampton, Southampton SO17 1BJ, UK}
\author{Przemys\l{}aw\,Oliwa}
\affiliation{Institute of Experimental Physics, Faculty of Physics, University of Warsaw, Poland}
\author{Rafa\l{}\,Mazur}
\author{Przemys\l{}aw\,Morawiak}
\author{Wiktor\,Piecek}
\affiliation{Institute of Applied Physics, Military University of Technology, Warsaw, Poland}
\author{Przemys\l{}aw\,Kula}
\affiliation{Institute of Chemistry, Military University of Technology, Warsaw, Poland}
\author{Pavlos\,G.\,Lagoudakis}
\affiliation{Skolkovo Institute of Science and Technology, Bolshoy Boulevard 30, bld. 1, Moscow, 121205, Russia}
\affiliation{Department of Physics and Astronomy, University of Southampton, Southampton SO17 1BJ, UK}
\author{Micha\l{}\,Matuszewski}
\affiliation{Institute of Physics, Polish Academy of Sciences, al.\,Lotnik\'{o}w 32/46, PL-02-668 Warsaw, Poland}
\author{Witold\,Bardyszewski}
\affiliation{Institute of Theoretical Physics, Faculty of Physics, University of Warsaw, Poland}
\author{Barbara\,Pi\k{e}tka}
\author{Jacek\,Szczytko}
\email{Jacek.Szczytko@fuw.edu.pl}
\affiliation{Institute of Experimental Physics, Faculty of Physics, University of Warsaw, Poland}

\begin{abstract}
Spin-orbit interactions which couple spin of a particle with its momentum degrees of freedom lie at the center of spintronic applications. Of special interest in semiconductor physics are Rashba and Dresselhaus spin-orbit coupling (SOC). When equal in strength, the Rashba and Dresselhaus fields result in SU(2) spin rotation symmetry and emergence of the persistent spin helix (PSH) only investigated for charge carriers in semiconductor quantum wells. Recently, a synthetic Rashba-Dresselhaus Hamiltonian was shown to describe cavity photons confined in a microcavity filled with optically anisotropic liquid crystal. In this work, we present a purely optical realisation of two types of spin patterns corresponding to PSH and the Stern-Gerlach experiment in such a cavity. We show how the symmetry of the Hamiltonian results in spatial oscillations of the spin orientation of photons travelling in the plane of the cavity.
\end{abstract}
\maketitle


The stability of a number of interesting phenomena in physical systems can be explained as a consequence of certain underlying symmetries that are robust to perturbations. Out of many examples in solid state physics, one that in the recent years has attracted a great deal of attention, is the persistent spin helix protected by an unusual SU(2) symmetry\,\cite{Bernevig_PRL2006}. It emerges in two dimensional semiconductors exhibiting Rashba and Dresselhaus SOC of equal magnitudes. The PSH is characterised by a spatially periodic spin texture of the SOC particles which become robust against spin-dependent scattering, with spin relaxation suppressed. So far, this effect has been experimentally demonstrated in many implementations \cite{Koralek_Nature2009, Walser_NatPhys2012, Kohda_2012, Ishihara2014, Sasaki_2014, Kohda_2017, Schliemann_RevModPhys2017, Passmann2018}. It was first observed through transient spin-grating spectroscopy \cite{Koralek_Nature2009}, then directly mapped in optical Kerr rotation experiment\,\cite{Walser_NatPhys2012}.

In the context of SOC of light \cite{Bliokh2015}, a rapidly growing field of research, there has been so-far no implementation of a photonic PSH to protect the polarization state of light. Today, SOC of light has led to remarkable results aiming to study well known concepts in solid state electronic systems but in a new optical context. This has contributed to development of exciting areas of study such as topological photonics \cite{Ozawa_RevModPhys2019}, nontrivial and singular polarization textures \cite{Leyder_NatPhys2007, Krol_Optica21}, valleytronics~\cite{Gong_Science2018, Lundt_NatNano2019}, and synthesising artificial gauge fields in photonic lattices~\cite{Whittaker_NatPhot2020}.

In this paper, we realise both a photonic PSH as well as optical Stern-Gerlach experiment using a liquid crystal–filled multimode cavity. By tuning two modes of opposite parity and polarization into resonance in this highly anisotropic cavity, their mixing becomes described by an effective equal Rasha-Dresselhaus SOC, resulting in a dispersion with strong valley polarization~\cite{Rechcinska_Science2019}. Our experimental observations and analytical calculations demonstrate that the strong polarization-valley coupling in this simple system directly results in the appearance of long-range polarization [or (pseudo)spin] textures of the in-plane traveling photons with potential application for valley-optronic devices \cite{Chen_NatPhoton2017,Sun2017,Dufferwiel_NatPhoton2017} alongside gapped Dirac materials. 



\begin{figure}[hbt]
    \centering
    \includegraphics{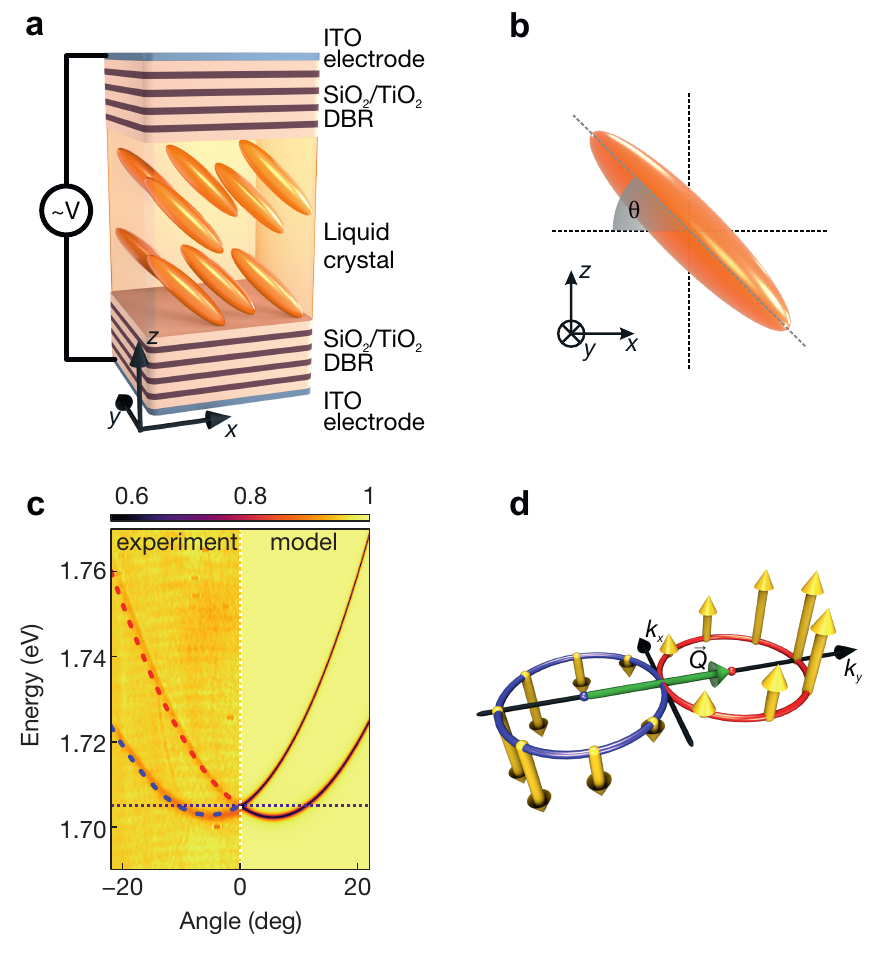} 
    \caption{\textbf{The effect of liquid crystal birefringence tuning.} \textbf{a}, Scheme of the cavity filled with nematic liquid crystal (LCMC). \textbf{b}, With external voltage applied to LCMC mean orientation of LC molecules tilt in \textit{x--z} plane. \textbf{c},\,Angle-resolved reflectance spectra in $y$ direction for Rashba-Dresselhaus resonance in LCMC: experiment and Berreman matrix simulation. Blue and red dashed lines mark fitted Rashba-Dresselhaus dispersion relation for spin-up and spin-down photons respectively. Dashed vertical line marks energy of the laser used for resonance transmission measurements. \textbf{d},\,Constant energy cross section through Rashba-Dresselhaus dispersion relation which consists of two circularly polarised circles off-centred by $\vec Q$ (green arrow). Yellow arrows denote the effective SOC magnetic field given by the last term in Eq.~\eqref{eq:Hexp}.}
    \label{im:expR}
\end{figure}

We use a microcavity filled with a liquid crystalline medium shown schematically in Fig.\,\ref{im:expR}a.  The cavity is based on two \ch{SiO2}/\ch{TiO2} distributed Bragg reflectors, with maximum reflectance at 1.65\,eV (750\,nm). Approximately 3.5\,$\upmu$m space between the DBRs is filled with a nematic liquid crystal of high birefringence ($\Delta n = 0.41$ \cite{Dabrowski2013, Miszczyk_LC2018}), which acts as an optically uniaxial medium inside a multimode cavity. By tuning an external voltage applied to transparent ITO electrodes, we can control the anisotropy direction in \textit{x--z} plane (Fig.\,\ref{im:expR}b), which changes the effective refractive index, thus cavity mode energy for light polarised in \textit{x} direction, whereas modes of perpendicular polarization are unaffected \cite{Lekenta_LSA2018}. If the refractive indices for perpendicular linear polarizations are different, then  degeneration of two modes with different numbers is possible. When two modes of opposite parities are degenerate they couple via Rashba-Dresselhaus SOC term and the dispersion can be described by an effective Hamiltonian written in the photon circular polarization basis (i.e., spin-up and spin-down states)\cite{Rechcinska_Science2019}:

\begin{equation} \label{eq:Hexp}
\hat{H}=\frac{\hbar^2{\vec k}^2}{2m}-2\alpha\hat{\sigma}_zk_y,
\end{equation}
which is the same as considered by Bernevig et al. in \cite{Bernevig_PRL2006}. Here, $ {\vec k}=(k_x, k_y)$ is the cavity in-plane momentum and $\hat{\sigma}_z$ is the third Pauli matrix. 


The Rashba-Dresselhaus SOC dispersion was achieved when 2.12\,V AC voltage applied to the sample. Dispersion relation of photons confined in the cavity can be mapped directly through angle-resolved reflectance spectra, as shown in  Fig.\,\ref{im:expR}c. The dispersion for wave vectors along the $y$ direction shows two off-centred spin-polarised parabolas (or valleys), where constant energy cross section  consists of two spin circles off-centred by momentum $\vec Q = 4m\alpha/\hbar^2\hat y$ (Fig.\,\ref{im:expR}d) with blue and red colours denoting spin-up and spin-down states, and yellow arrows indicating the effective momentum-dependent out-of-plane magnetic field. Fitting Eq.~\eqref{eq:Hexp} to the dispersion gives a Rashba parameter $\alpha = 2.8$\,meV$\cdot\upmu$m and effective mass $m = 1.1\cdot10^{-5} m_{\rm e}$ where $m_e$ is the free electron rest mass. The angle-resolved reflectance spectra was also simulated numerically using the Berreman method \cite{Berreman_1972,Schubert_PRB1996}. Results of such simulations for LC layer width of 3.1\,$\upmu$m with molecules rotated by $\theta=53.3^\circ$ are compared with experiment in Fig.\,\ref{im:expR}c.

\begin{figure*}[hbt]
    \centering
    \includegraphics{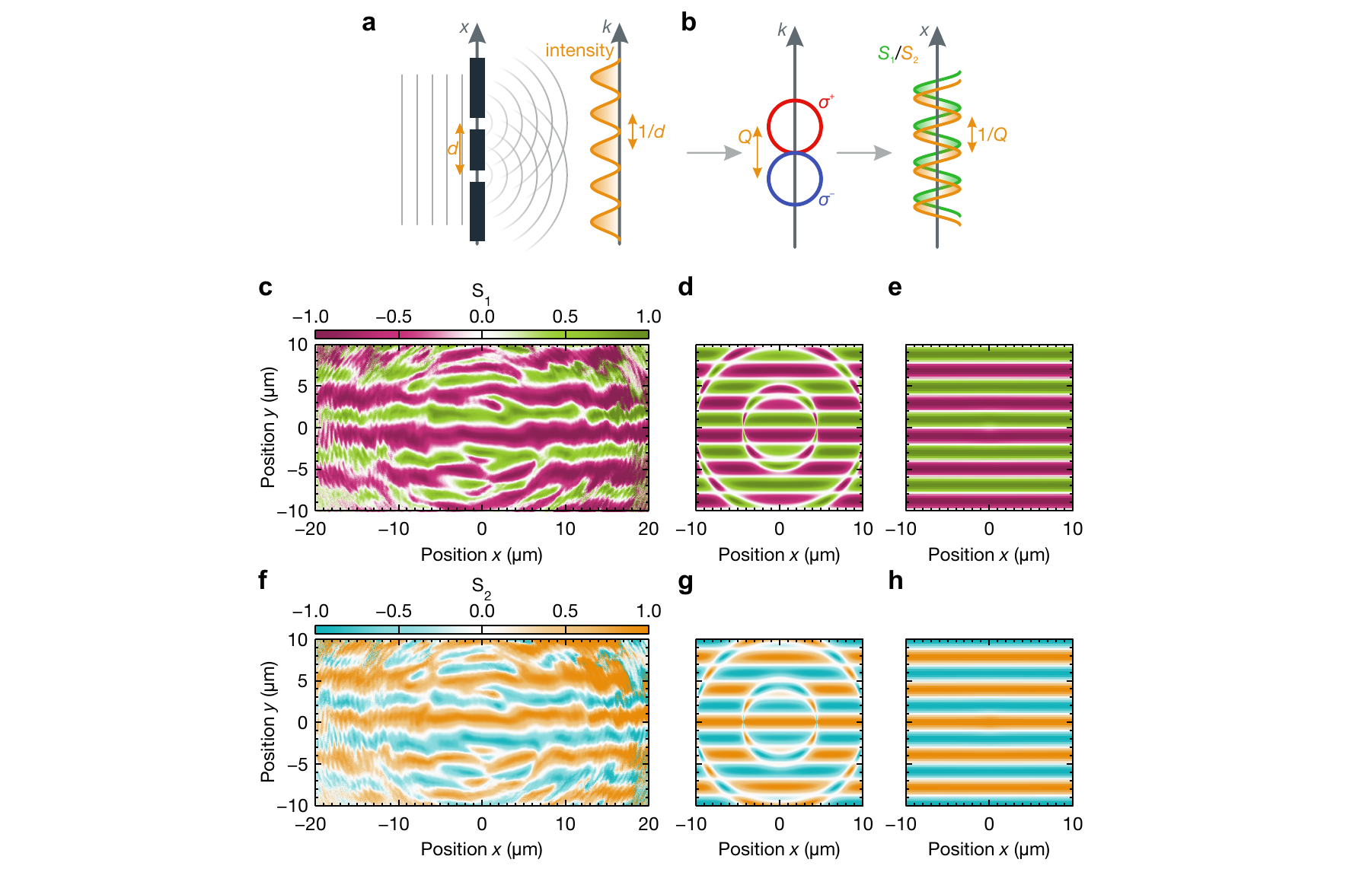}
    \caption{\textbf{All-optical persistent spin helix state.}  \textbf{a}, Schematic of Young's double slit experiment. \textbf{b}, Schematic of the reciprocal spin-valley interference experiment. \textbf{c,f,} Experimental spatial $S_1$ and $S_2$ Stokes parameters of the light transmitted through the cavity under tightly focused monochromatic antidiagonally polarised incident light. \textbf{d,g,} Corresponding patterns calculated from the analytical formula~\eqref{eq.Stokes}, and \textbf{e,h,} from Schr\"{o}dinger equation simulation of Eq.~\eqref{eq:Hexp}.}
    \label{im:modDCP}
\end{figure*}

Interestingly, one can consider the equal Rashba-Dresselhaus SOC system as a spin realisation of Young's double slit experiment in a reciprocal space. The role of the two slits discriminating the position of an incident scalar plane wave in the real space---resulting in the well known double-slit spatial interference pattern in the far field---is instead played by the two cavity dispersion spin valleys in reciprocal space discriminating the momenta of light through an optical mode (Fig.\,\ref{im:expR}d). In the double-slit experiment a plane wave passing through two slits (separated by distance $d$) produces spherical waves that interfere, giving an image of the intensity oscillating in space with a period proportional to $1/d$ (Fig.\,\ref{im:modDCP}a). Analogously, for a homogeneous occupation of the two isoenergy spin circles in the reciprocal space, separated by the vector $\vec{Q}$, we would obtain in real space a polarization interference image producing the persistent spin helix with a period proportional to $L=2\pi/Q$ where $Q=|\vec{Q}|$. In other words, in the Young double slit experiment, one obtains in the far field the Fourier transform of the two slits, which is a periodic interference pattern. Here, we obtain in the near field the Fourier transform of the the two polarised circles in momentum space, which is a polarization interference pattern.  Based on the fitted dispersion in Fig.\,\ref{im:expR}c this leads to a helix period of $L=3.8$\,$\upmu$m.

This change of the polarization with propagation of photons in the plane of cavity is observed experimentally in our system. An incident linearly antidiagonally polarised laser beam is tightly focused with a microscope objective on the sample, providing homogeneous occupation of photons on both spin circles in reciprocal space. The laser energy is set resonant with the cavity modes at normal incidence as marked by dashed horizontal line in Fig.\,\ref{im:expR}c. Transmitted light is collected by another microscope objective, polarization resolved  and imaged on a CCD camera. This allows us to map out the spatial distributions of the $S_1=(I_X - I_Y)/(I_X + I_Y)$ and $S_2=(I_d - I_a)/(I_d + I_a)$ and $S_3=(I_{\sigma^+} - I_{\sigma^-})/(I_{\sigma^+} + I_{\sigma^-})$ Stokes parameters, corresponding to intensities of horizontal ($I_X$), vertical ($I_Y$), diagonal ($I_d$), antidiagonal ($I_a$), right-hand circular ($I_{\sigma^+}$) and left-hand circular ($I_{\sigma^-}$)  polarised light.

The measured $S_1$ and $S_2$ parameters are plotted in Fig.\,\ref{im:modDCP}c,f, which clearly show periodic oscillations with $\pi/2$ phase shift between the two Stokes parameters.
Spatial period of the oscillations is estimated as $L=4.7$\,$\upmu$m. The phase of the PSH depends on the polarization of incident light, as described in the Supplemental Material~\cite{SI}.

 As mentioned above, this result can be understood as a consequence of an interference process between spins in different momentum valleys (i.e., valley polarization).
  Assuming that the cavity extends infinitely in the $x$--$y$ plane with the two almost perfect mirror planes separated by distance $L$, we can represent the modal electric fields inside the cavity, corresponding to the eigenvalues $\varepsilon^\pm(\vec k) = \frac{\hbar^2}{2m}\left(\vec k \mp \frac{\vec Q}{2}\right)^2 - \frac{2m\alpha^2}{\hbar^2}$ of the Hamiltonian (\ref{eq:Hexp}) by the plane waves:
  \begin{equation}
    \label{eq:wav}
    \Psi^\pm_{\vec k}(\vec r,z) =
\left[\begin{array}{l}E_X\\ \\E_Y\end{array}\right]_\pm = e^{i\vec k\cdot\vec r} 
\left[\begin{array}{l}\frac{1}{n_x}\sin\frac{M\pi z}{L}\\ \\\frac{\mp i }{n_y}\sin\frac{N\pi z}{L} \end{array}\right].
\end{equation}
Here $n_x$ and $n_y$  represent the  refractive indices for linear polarizations in $x$ and $y$ directions, respectively, and $M,N$ denote the degenerate longitudinal cavity mode numbers of opposite parity, i.e., $M = N \pm 1$. The electric field of  the modes $\Psi^\pm_{\vec k}$ in vicinity of the mirrors is right/left hand circularly polarised with respect to the normal vector pointing outside the cavity plane. So a plus sign corresponds to an outgoing wave with
$\vec\sigma_+ = \frac{1}{\sqrt{2}}\left[1,i\right]^T$ polarization vector and the minus sign corresponds to an outgoing wave with the
$\vec\sigma_-=\frac{1}{\sqrt{2}}\left[1, -i\right]^T$ polarization vector on either side of the cavity. The planar symmetry of the cavity implies that incident electromagnetic waves of energy $\varepsilon$ and momentum $\vec{k}$ will excite modes $\varepsilon^\pm(\vec{k})$.
A linearly polarised incident plane wave with the polarization angle $\Theta = \frac{1}{2} \tan^{-1}{(S_2/S_1)}$ with respect to the $x-$axis will excite either $e^{+i\Theta}\Psi^+$ or $e^{-i\Theta}\Psi^-$ waves inside the cavity, 
provided $\vec k$ belongs to the red or blue circle in Fig.\,\ref{im:expR}d. Therefore the transmitted light will be either right-hand or left-hand circularly polarised.  If, however, the surface of the microcavity is illuminated with a focused coherent beam at normal incidence then the entire isoenergy momentum spin circles are excited and a specific polarization pattern determined by $\Theta$, as a manifestation of the optical PSH, will emerge. Let $F$ denote the distance between the focus of the incident beam and the illuminated surface. The electric field at this surface is a combination of plane waves (up to a common factor):
    \begin{equation} \label{eq.Ein}
      \vec E_\text{in} =  \int \frac{d^2k}{(2\pi )^2}e^{i\vec k\cdot \vec r}
      e^{iR^2k^2 } \left[\begin{array}{l} \cos\Theta\\ \sin\Theta\end{array}\right],
    \end{equation}
    where $R = \sqrt{F/2k_0}$ and $k_0$ is equal to the light wavenumber in the vacuum.
    Taking into account that each of those plane waves couples either to the field
    $ e^{+i\Theta}\Psi^+$ or  $ e^{-i\Theta}\Psi^-$ inside the cavity
    and performing the integral with respect to $\vec k$ one arrives at the final expression for the electric field of the transmitted wave at the opposite surface. Denoting $\phi_\pm \triangleq {\cal J}_0(\sqrt{\frac{2m}{\hbar^2}(E+\frac{2m\alpha^2}{\hbar^2})}|\vec r \pm R^2\vec Q|)$,  where $  {\cal J}_0$ is the Bessel function of zeroth order, we obtain
    \begin{equation} \label{eq.nonsep}
      \vec E_\text{ out } \sim  e^{i\frac{\vec Q}{2}\cdot\vec r}\phi_+(\vec r)\vec\sigma_+
      +  e^{-i\frac{\vec Q}{2}\cdot\vec r}\phi_-(\vec r)\vec\sigma_-.
    \end{equation}
This expression exemplifies the so-called {\it classical entanglement}, while some authors find such terminology misleading \cite{Karimi_Science2015}, between the planar position and the polarization of the electric field \cite{Forbes_AdvInOpt2019}. Or alternatively, we have created an inseparable state between the valley $\pm \vec{Q}/2$ and polarization $\vec{\sigma}_\pm$ degrees of freedom (DOF) of the cavity photons. The spatial polarization pattern of the solution~\eqref{eq.nonsep} can be represented by the components of the normalised Stokes vector,
 \begin{equation} \label{eq.Stokes}
   \begin{aligned}
     S_1 &= \frac{2 \phi_+(\vec r)\phi_-(\vec r)}{\phi_+(\vec r)^2+\phi_-(\vec r)^2} \cos(\vec Q\cdot\vec r+2\Theta), \\
     S_2 &= -\frac{2 \phi_+(\vec r)\phi_-(\vec r)}{\phi_+(\vec r)^2+\phi_-(\vec r)^2} \sin(\vec Q\cdot\vec r+2\Theta), \\
     S_3 &= \frac{\phi_-(\vec r)^2-\phi_+ (\vec r)^2}{\phi_+(\vec r)^2+\phi_-(\vec r)^2}.
   \end{aligned}
   \end{equation}
The analytical results \eqref{eq.Stokes} are plotted in Fig.\,\ref{im:modDCP}d,g for $\Theta=-\pi/4$, which agree with the experimental data obtained using antidiagonally polarised irradiation. More strict analytical calculations taking into account the Gaussian profile of the focused beam does not change significantly above conclusions. Additionally, we perform Schr\"{o}dinger equation simulations $i \hbar \partial_t \Psi = \hat{H}(-i\nabla) \Psi + \vec{f}(\vec{r})$ driven by a normally incident antidiagonally polarised narrow Gaussian field $\vec{f}(\vec{r}) = e^{-r^2/(2w^2)} [e^{i\pi/4}, -e^{-i\pi/4}]^T$ under open boundary conditions~\cite{SI}. The results are shown in Fig.\,\ref{im:modDCP}e,h in good agreement with experiment and analytical theory. 

Our results open several new exciting perspectives in photonics. The so-called classically entangled DOF between the cavity dispersion valleys and the photon spin offer a transparent and easy way to create inseparable photonic states~\cite{Kagalwala_NatPho2013}. Indeed, by controlling the polarization of the incident optical beam defined as $\vec{\sigma}_\text{in} = \beta_+ \vec{\sigma}_+ +  \beta_- \vec{\sigma}_-$, where $|\beta_+|^2 + |\beta_-|^2 = 1$, one can write Eq.~\eqref{eq.nonsep} as a function of two control variables,
\begin{equation} \label{eq.nonsep2}
   \vec E_\text{ out }(\beta_\pm) \sim \beta_+ e^{i\frac{\vec Q}{2}\cdot\vec r}\phi_+(\vec r)\vec\sigma_+
      +  \beta_- e^{-i\frac{\vec Q}{2}\cdot\vec r}\phi_-(\vec r)\vec\sigma_-
\end{equation}
It then becomes apparent that a family of inseparable states exist which satisfy $|\beta_+| = |\beta_-|$ and $\beta_+/\beta_- = e^{2 i \Theta}$. Similar to a Michelson interferometer, $\Theta$ is an effective "path-difference" variable which uniquely determines the location of the $S_{1,2}(\vec{r})$ interference minima and maxima in the PSH [see SI and Eq.~\eqref{eq.Stokes}]. This reflects the fact that for an inseparable state, any effects on one DOF (e.g., spin) will have measurable outcome in the other DOF (e.g., momentum valley) with exciting potential in optical metrology that benefits from parallelised DOF measurements~\cite{Toppel_IOP2014}.

In the case of a tightly focused pump, as considered above, one has $\phi_+(\vec{r}) \simeq \phi_-(\vec{r})$ and the amount of nonseparability can be quantified as a global parameter $C = 2|\beta_+ \beta_-|$ where $0 \leq C \leq 1$~\cite{Korolkova_RPP2019}. When $C=1$ the system is maximally inseparable whereas when $C = 0$ the spin and valley degrees of freedom are completely separable. In analogy to the double slit experiment, where one can control the intensities of light passing through each slit affecting the interference pattern, setting $C<1$ describes different amounts of photons  in the two circularly polarised component of light resulting in imbalanced occupation of the two valleys. In the more general case where $\phi_+(\vec{r}) \neq \phi_-(\vec{r})$ (causing the interference circles in the $S_{1,2}$ in Fig.~\ref{im:modDCP}d,g) the amount of inseparability is no longer a global quantity and depends on the spatial coordinate $\vec{r}$. In this case the $S_3(\vec{r})$ Stokes parameter becomes a useful measure on the amount of inseparability in space written as $C^2 + S_3^2=1$, 
which is an analog of complimentarity proposed by Eberly et al. \cite{Eberly_2016}.

\begin{figure}[hbt]
    \centering
    \includegraphics{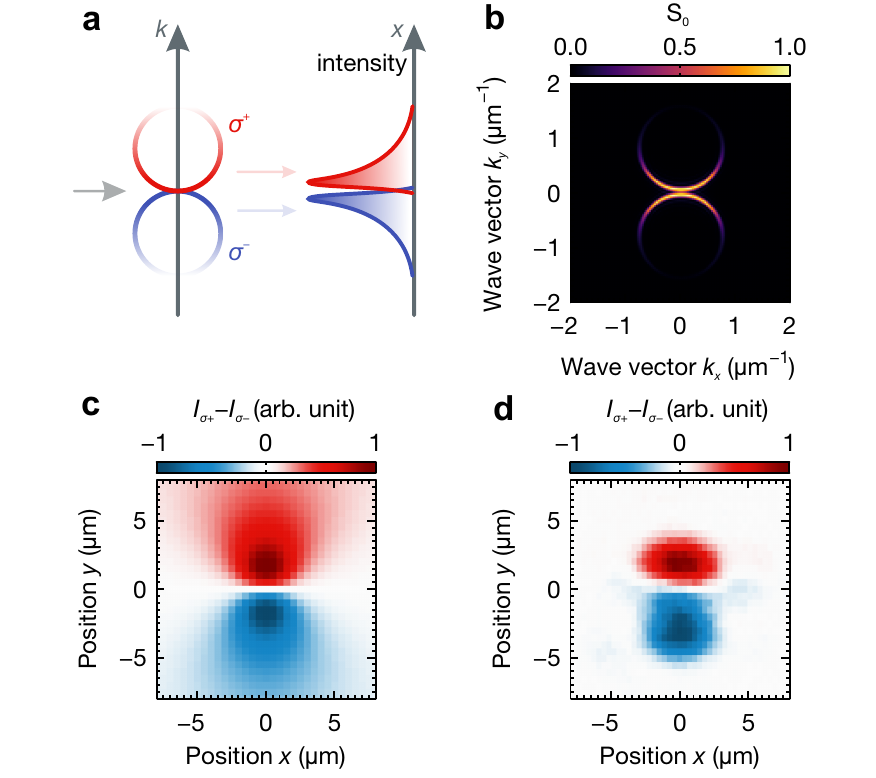}
    \caption{\textbf{Photonic Stern-Gerlach experiment.} \textbf{a,}\,Schematic of the reciprocal Stern-Gerlach experiment where the isoenergy circles of both valleys are partially populated resulting in separation of circular polarization along the $y$ direction. \textbf{b,} The intensity of light in reciprocal space ($|\tilde{\Psi}(\vec{k})|^2$) under linearly polarised wide Gaussian driving field $\vec{f}$. \textbf{c,} Schr\"{o}dinger simulations of the difference between transmitted light intensities in $\sigma^+$ and $\sigma^-$ polarizations. \textbf{d,}\,Experimental difference between light intensities in $\sigma^+$ and $\sigma^-$ polarizations obtained from real space tomography under incoherent light transmission through the cavity using a wide Gaussian beam.} 
    \label{im:expDCPwl}
\end{figure}

Moreover, scrutinising the $S_3$ parameter we can demonstrate analogy to an optical Stern-Gerlach experiment~\cite{Karpa_NatPhys2006} using our system. The effective magnetic field of the equal Rasha-Dresselhaus SOC causes a spin-selective deflection of the cavity photons along the two opposite directions in the cavity plane defined by the valleys location $\pm \vec{Q}/2$ (Fig.\,\ref{im:expR}d). This deflection appears due to the different Bessel solutions for $\vec{\sigma}_+$ and $\vec{\sigma}_-$ which are shifted due to the anisotropy of the dispersion. This can be evidenced for the linearly polarised case $C = 1$. For non-homogeneous occupation of dispersion cavity valleys (Fig.~\ref{im:expDCPwl}b)  we observe that $S_3(x,y) = -S_3(x,-y)$ as indicated theoretically in Fig.~\ref{im:expDCPwl}c and experimentally in Fig.~\ref{im:expDCPwl}d. Such non-homogeneous occupation of the valleys can be achieved by using a broad normal incident excitation beam which only excites a locality in reciprocal space around $k=0$. In the experiment presented in Fig.~\ref{im:expDCPwl}d we used linearly polarised (diagonal) light from broadband halogen lamp  transmitted through the sample by the same optical system (see Methods). 
Our results therefore open a new method to design an optical Stern-Gerlach experiment in the classical optics regime. The notable difference between our setup and the actual Stern-Gerlach is that there is no constant force acting on the photon pseudospins but rather, they obtain constant group velocities in opposite directions in the cavity due to the effective magnetic force gradient (yellow arrows in Fig.~\ref{eq:Hexp}d). This result should not be confused with valley selective circular dichroism~\cite{Cao_NatComm2012} as there are no absorption processes involved here, or the much weaker spin Hall effect of light coming from geometric phases~\cite{Hosten_SCience2008}, or circular birefringence in chiral materials like the Fresnel triprism~\cite{Arteaga_OptExp2019} since our system is achiral (i.e., the $x$-$z$ plane is a plane of symmetry). A future path of investigation can then involve a quantum Stern-Gerlach experiment operating in the single photon regime. Finally, implementation of multiple spatially separated incident optical beams given by their central coordinates $\{\vec{r}_1, \vec{r}_2,\dots\}$ introduces an additional external DOF that defines the beams location. Future photonic multi-DOF experiments combining coordinate (i.e., incident beam central location), valley, and spin DOFs can be engineered, producing more complex in-plane polarization patterns as a result of interference.

In summary, we have investigated a liquid crystal filled optical cavity with equal Rashba and Dresselhaus SOC contributions leading to SU(2) spin rotation symmetry \cite{Bernevig_PRL2006}. An illustration of such symmetry is the emergence of the persistent spin helix, previously shown only in electronic systems \cite{Koralek_Nature2009,Walser_NatPhys2012}. This effect allows for fine-control over the spatial polarization state of the in-plane cavity photons on a micrometer scale.
We have interpreted our observations in the framework of inseparable degrees of freedom played by the Rashba-Dresselhaus momentum valleys and the photon spin. This has allowed us to establish a reciprocal optical version of the famous Young's double slit experiment and the Stern-Gerlach experiment. Due to the compact design LCMC can be easily integrated with optoelectronics devices, giving the perspective of simulating complex spin systems, studying effects that are difficult to control in condensed matter systems, and developing photonic valleytronic devices in a relatively simple setting at room temperature.


\textbf{Methods}

For transmission measurements laser light from Ti:Sapphire laser with energy 1.705\,eV is focused with 100$\times$ microscope objective and collected by another objective with 50$\times$ magnification. Both objectives have numerical aperture equal to 0.55.  Incident light polarization is linearly polarised in antidiagonal direction.

Data in Fig.~\ref{im:expDCPwl}d was obtained by polarization-resolved tomography. Linearly polarised (in diagonal direction) light from broadband halogen lamp was transmitted through the sample by the same optical system. Transmitted light was imaged by lens with 400\,mm focal length on entrance slit of a monochromator equipped with a CCD camera. Tomography was obtained by motorised movement of the imaging lens in direction perpendicular to the slit. Fig.~\ref{im:expDCPwl}d presents cross section at the energy of Rashba-Deresselhaus resonance.

\textbf{Data availability}

The data that support the plots within this paper and other findings of this study are available from the corresponding authors upon reasonable request.

\bibliography{bib}

\textbf{Acknowledgements}

This work was supported by the Ministry of Higher Education, Poland, under project “Diamentowy Grant”: 0005/DIA/2016/45, the National Science Centre grant 2019/35/B/ST3/04147 and 2019/33/B/ST5/02658, and the Ministry of National Defense Republic of Poland Program -- Research Grant MUT Project 13–995 and MUT University grant (UGB) for the Laboratory of Crystals Physics and Technology for year 2021. H.S. and P.G.L. acknowledge the support of the UK’s Engineering and Physical Sciences Research Council (grant EP/M025330/1 on Hybrid Polaritonics), the support of the RFBR project No. 20-52-12026 (jointly with DFG) and No. 20-02-00919, and the European Union’s Horizon 2020 program, through a FET Open research and innovation action under the grant agreement No. 899141 (PoLLoC) and No. 964770 (TopoLight).

\textbf{Author contributions}

M.K., K.R., P.O. and S.P. performed the experiments under the guidance of B.P. and J.S. P.K. synthesized liquid crystal. R.M., P.M., and W.P. constructed and fabricated the LC microcavity. W.B., P.O.,  M.M.  worked on the theoretical description of optical PSH and H.S. of optical Stern-Gerlach.  J.S., B.P., W.P., and P.G.L. supervised the project. K.R., M.K., H.S. and J.S. wrote the manuscript with input from all other authors.

\textbf{Competing interests}

The authors declare no competing interests.

\textbf{Additional information}

Supplementary information is available for this paper.

Correspondence and requests for materials should be addressed to J.S.


\newpage

\end{document}